\documentclass[referee, useAMS, usenatbib, amstex, amssymb]{mn2e}
\usepackage{psfig}
\usepackage{graphicx}
\usepackage{epsf}
\usepackage{bm}
\usepackage{lscape}

\title [Radial Metallicity Gradient from RAVE DR3]
{Local stellar kinematics from RAVE data:
II. Radial Metallicity Gradient}

\author[Co\c skuno\u glu et al.]
       {B. Co\c skuno\u glu $^{1}$\thanks{E-mail: basarc@istanbul.edu.tr},
S. Ak$^{1}$, S. Bilir${^1}$, S. Karaali$^{1}$, \"O. \"Onal$^{1}$, E. Yaz$^{1}$,
\newauthor
 G. Gilmore$^{2,3}$, G. M. Seabroke$^{4}$
\\
  $^1$Istanbul University, Science Faculty, Department of Astronomy and Space
Sciences, 34119, University-Istanbul, Turkey\\
  $^2$Institute of Astronomy, Madingley Road, Cambridge, CB3 OHA, UK\\
  $^3$Astronomy Department, Faculty of Science, King Abdulaziz University, P.O.
Box 80203, Jeddah 21589, Saudi Arabia\\
  $^4$Mullard Space Science Laboratory, University College London, Hombury St Mary, 
Dorking, RH5 6NT, UK\\}

\date{}

\pagerange{\pageref{firstpage}--\pageref{lastpage}} \pubyear{}

\begin{document}

\maketitle

\label{firstpage}
\begin{abstract}
We investigate radial metallicity gradients for a sample of dwarf
stars from the RAdial Velocity Experiment (RAVE) Data Release 3 (DR3). 
We select a total of approximately 17000
F-type and G-type dwarfs using selection of colour, $\log g$, and
uncertainty on derived space motion, calculate for each star a
probabilistic (kinematic) population assignment to thick or thin disc
using space motion, and additionally another (dynamical) assignment using
stellar vertical orbital eccentricity. We additionally subsample by
colour, to provide samples biased to young thin disc and to older
thin disc stars.  We derive a metallicity gradient as a function of
Galactocentric radial distance, i.e. $d[M/H]/dR_m=-0.051\pm0.005$ dex
kpc$^{-1}$, for the youngest sample, F-type stars with vertical orbital 
eccentricities $e_V\leq0.04$. Samples biased to older thin disc stars show
systematically shallower abundance gradients.
\end{abstract}


\section{Introduction}

Abundance gradients in galactic discs are an important constraint on
the star formation and interstellar medium history and the possible
secular evolution of stars post-formation in those discs. In the Milky
Way Galaxy there is extensive information establishing a radial
gradient in young stars and in the interstellar medium
\citep*{Shaver83, Luck06, Luck11b}. Values typically derived are
$d[Fe/H]/dR\rm_G=-0.06\pm0.01$ dex kpc$^{-1}$, within 2-3 kpc of the
Sun. Much effort to search for local abundance variations, and
abundance variations with azimuth, show little if any detectable
variation in young systems \citep{Luck11a}, limiting the importance of
radial gas flows.

Quantifying the abundance distribution functions and their radial and
vertical gradients in both the thin and thick discs can be achieved
using stellar abundances, especially those from major surveys such as
RAdial Velocity Experiment \citep[RAVE,][]{Steinmetz06, Zwitter08, Siebert11}. 
Several recent analyses of RAVE data have studied the 
metallicity and kinematics of the Galactic discs distribution functions from
RAVE data \citep{Burnett11, Karatas11, Ruchti11, Wilson11}.

Even more fundamental is the variation of the abundance gradient with
time. Some information is available, especially from open star
clusters and field stars. The evidence here suggests the gradient
flattens beyond about 12 kpc Galactocentric radius \citep*{Carney05, Yong05}, as
is seen in other spiral galaxies \citep*{Worthey05, Vlajic09}. At very
outer radii the situation may be more complex.

The observational situation however can be improved.
Extant data suggest vertical metallicity gradients in the
$-0.4<d[M/H]/dz<-0.2$ dex kpc$^{-1}$ range for relatively small
distances from the Galactic plane, i.e. $z<4$ kpc \citep*{Trefzger95, 
Karaali03, Du04, Ak07a}. For intermediate $z$ distances, where the
thick disc is dominant, the vertical metallicity gradient is low,
$d[M/H]/dz=-0.07$ dex kpc$^{-1}$ and the radial gradient is only
marginal, $-0.02\leq d[M/H]/dz\leq0$ dex kpc$^{-1}$ \citep*{Rong01}.
There is some evidence that metallicity gradients for
relatively short vertical distances, $z<2.5$ kpc, show systematic
fluctuations with Galactic longitude, similar to those of the
thick-disc scaleheight, which may be interpreted as a
common flare effect of the disc \citep{Ak07b, Bilir08, Yaz10}.

These vertical gradients are perhaps a convolution of a time-dependent abundance 
and the well-established stellar age-velocity dispersion relation. 
Alternatively, they may be more consistent with two independent, 
rather narrow metallicity distributions, one thin disc, one thick disc, 
with the apparent gradient being simply a reflection of the two different 
scale heights \citep{GW85, Burnett11}.

In this study we analyse stellar abundance gradients from RAVE data,
using colour bins as a proxy for mean age, to test for time-dependent
effects. RAVE is a multi-fibre spectroscopic astronomical survey of
stars in the Milky Way, which covers just over half of the Southern
hemisphere, using the 1.2 m UK Schmidt Telescope of the Australian
Astronomical Observatory (AAO). RAVE's primary aim is to derive the
radial velocity of stars from the observed spectra for solar
neighbourhood stars. Additional information is also derived, such as
photometric parallax and stellar atmospheric parameters,
i.e. effective temperature, surface gravity, metallicity and elemental
abundance data. This information is important in calculating
metallicity gradients, which provides data about the formation and
evolution of the Galaxy. As the data were obtained from solar neighbourhood stars, we
have limitations to distance and range of metallicity. However,
the metallicity measurements are of high internal accuracy which is an
advantage for our work. A radial metallicity gradient of -0.04 dex
kpc$^{-1}$, based on calibrated metallicities from the RAVE DR2 data
has already appeared in the literature \citep{Karatas11}. However,
this metallicity gradient covers all spectral types, and thus a wide
age range, which is a disadvantage for a sample of stars with distance
and metallicity range restrictions. Hence, in this study we will use
only the data for the F and G spectral type stars. We test for
different metallicity gradients for different spectral types.

The structure of this paper is: Data selection is described in Section
2; calculated space velocities and orbits of star samples and
population analysis are described in Sections 3 and 4,
respectively. Results are given in Section 5 and a summary and
conclusion are presented in Section 6.

\section{Data}
The data were selected from the third data release of RAVE 
\citep[DR3 -][]{Siebert11}. RAVE DR3 reports 83072 radial velocity
measurements for stars with $9\leq I\leq12$. This release also
provides stellar atmospheric parameters for 41672 spectra representing
39833 individual stars \citep{Siebert11}. The accuracy of the radial
velocities is high, marginally improved with DR3: the distribution of
internal errors in the radial velocities has a mode of 0.8 km s$^{-1}$
and a median of 1.2 km s$^{-1}$, while 95 per cent of the sample has an
internal error smaller than 5 km s$^{-1}$. The uncertainties for the
stellar atmospheric parameters are: 250 K for effective temperature 
$T_{eff}$, 0.43 dex for surface gravity $\log g$ and 0.2 dex for $[M/H]$. 
While RAVE supports a variety of chemical abundance scales, 
we use here just the public DR3 values. Since anticipated gradients are small, 
this provides a well-defined set of parameters for analysis.

A surface gravity constraint $4<\log g\leq5$ was applied to the 83072
star sample to obtain a homogeneous dwarf-star subsample with
accurate data. Distances of the subsample stars were obtained using
the absolute magnitude calibration of \cite{Bilir08}, whereas the
reddening values were obtained iteratively, following published
methodology \citep[for more detailed information regarding the 
iterations see][and the references therein]{Coskunoglu11}.

Our aim here is to take advantage of RAVE's focus on high Galactic
latitude to isolate, statistically, samples dominated by thin disc and
by thick disc stars. Thus, after dereddening, stars were
split into F and G spectral type subsamples using \cite{Straizys09}'s
criteria, i.e. $0.09<(J-H)_0\leq0.29$ and $0.29<(J-H)_0\leq0.39$
leaving us with 17768 stars. Out of these 17768 stars 10661 are F- and
7107 are G-type dwarfs, while the sample median distances are 326 and 272 pc,
respectively.

\section{Space Velocities and Orbits}
We combined the distances estimated in Section 2 with RAVE kinematics
and the available proper motions, applying the (standard) algorithms
and the transformation matrices of \cite{Johnson87} to obtain their
Galactic space velocity components ($U$, $V$, $W$). In the
calculations epoch J2000 was adopted as described in the
International Celestial Reference System (ICRS) of the {\em Hipparcos}
and {\em Tycho-2} Catalogues \citep{ESA97}. The transformation
matrices use the notation of a right handed system. Hence, $U$, $V$
and $W$ are the components of a velocity vector of a star with respect
to the Sun, where $U$ is positive towards the Galactic centre
($l=0^{\circ}$, $b=0^{\circ}$), $V$ is positive in the direction of
Galactic rotation ($l=90^{\circ}$, $b=0^{\circ}$) and $W$ is positive
towards the North Galactic Pole ($b=90^{\circ}$).

Correction for differential Galactic rotation is necessary for
accurate determination of $U$, $V$ and $W$ velocity components. The
effect is proportional to the projection of the distance to the stars
onto the Galactic plane, i.e. the $W$ velocity component is not
affected by Galactic differential rotation \citep{Mihalas81}. We
applied the procedure of \cite{Mihalas81} to the distribution of the
sample stars and estimated the first order Galactic differential
rotation corrections for $U$ and $V$ velocity components of the sample
stars. The range of these corrections is $-25.71<dU<16.92$ and
$-1.57<dV<2.29$ km s$^{-1}$ for $U$ and $V$, respectively. As
expected, $U$ is affected more than the $V$ component. Also, the high
values for the $U$ component show that corrections for differential
Galactic rotation can not be ignored.

The uncertainty of the space velocity components $U_{err}$, $V_{err}$
and $W_{err}$ were computed by propagating the uncertainties of the
proper motions, distances and radial velocities, again using a
(standard) algorithm by \cite{Johnson87}. Then, the error for the
total space motion of a star follows from the equation:

\begin{equation}
S_{err}^{2}=U_{err}^{2}+V_{err}^{2}+W_{err}^{2}.
\end{equation}
The median and standard deviation for space velocity errors are 
\~S$_{err}=8.01$ km s$^{-1}$ and $s=16.99$ km s$^{-1}$, respectively. We
now remove the most discrepant data from the analysis, knowing that
outliers in a survey such as this will preferentially include stars
which are systematically mis-analysed binaries, etc. Astrophysical
parameters for such stars are also likely to be relatively
unreliable. Thus, we omit stars with errors that deviate by more 
than the sum of the standard error and the standard deviation, 
i.e. $S_{err}>$25 km s$^{-1}$. This removes 760 stars, $\sim$4.3 
per cent of the sample. Thus, our sample was reduced to 17008 stars, 
those with more robust space velocity components. After applying 
this constraint, the median values and the standard deviations for 
the velocity components were reduced to (\~U$_{err}$, \~V$_{err}$, 
\~W$_{err}$)=($3.93\pm3.10$, $3.59\pm2.79$, $3.16\pm2.75$) km s$^{-1}$.

To complement the chemical abundance data, accurate kinematic data
have been obtained and used to calculate individual Galactic orbital
parameters for all stars. In order to calculate those parameters we
used standard gravitational potentials well-described in the
literature \citep*{Miyamoto75, Hernquist90, Johnston95, Dinescu99} to
estimate orbital elements of each of the sample stars. The orbital
elements for a star used in our work are the mean of the corresponding
orbital elements calculated over 15 orbital periods of that specific
star. The orbital integration typically corresponds to 3 Gyr, and is
sufficient to evaluate the orbital elements of solar
neighbourhood stars, most of which have orbital periods below 250
Myr.

Solar neighbourhood velocity space includes well-established
substructures that resemble classic moving groups or stellar streams
\citep*{Dehnen98, Skuljan99, Nordstrom04}. \cite*{Famaey05, Famaey08} and
\cite{Pompeia11} show that, although these streams include clusters,
after which they are named, and evaporated remnants from these
clusters, the majority of stars in these streams are not coeval but
include stars of different ages, not necessarily born in the same
place nor at the same time. They argue these streams are dynamical
(resonant) in origin, probably related to dynamical perturbations by
transient spiral waves \citep*{DeSimone04}, which radially migrate
stars to specific regions of the $UV$-plane. Stars in a dynamical
stream just share a common velocity vector at this particular epoch.
These authors further point out the obvious and important point that
dynamical streams are kinematically young and so integrating backwards
in a smooth stationary axisymmetric potential the orbits of the stars
belonging to these streams is non-physical. Does this fundamentally
invalidate our calculations?

\cite{Famaey05} assigned probabilities for each star in their sample
to belong to different kinematic groups. They found that stars not
belonging to dynamical streams (young giants and the smooth
background) make up the majority (70 per cent) of their sample.
\cite{Seabroke08}'s fig. 10 illustrates that the majority of
\cite{Famaey05}'s stars are within $\pm250$ pc of the Galactic plane,
with the distribution centred on the Galactic plane. Our RAVE sample
of F-G dwarfs are at similar line-of-sight distances from the Sun as
the \cite{Famaey05}'s K-M giants (200-400 pc). However,
\cite{Siebert11}'s fig. 15 shows that RAVE stars are selected to avoid
the Galactic plane ($|b|>10^{\circ}$). This means our RAVE sample
will include stars further from the Galactic plane than the
\cite{Famaey05} sample. Dynamical perturbations by transient spiral
waves are strongest closest to the Galactic plane so there will be
fewer dynamical stream stars in our RAVE sample. While we could in
principle assign our stars to different kinematic groups, compared to
\cite{Famaey05}'s, it is probable that our sample has more than the
local 70 per cent of stars whose orbital parameters can adequately be
determined from the static Milky Way disc potential and not influenced
by transient spiral waves. \cite{Famaey05}'s space velocities are also
much more accurate than our RAVE values, so considering all factors,
we have not attempted to remove dynamical stream stars from our
sample. Contamination is unlikely to affect our statistical
results.

To determine a possible orbit, we first perform test-particle
integration in a Milky Way potential which consists of a logarithmic
halo of the form
\begin{eqnarray}
  \Phi_{\rm halo}(r)=v_{0}^{2} \ln \left(1+\frac{r^2}{d^2}\right),
\end{eqnarray}
with $v_{0}=186$ km s$^{-1}$ and $d=12$ kpc. The disc is represented
by a Miyamoto-Nagai potential:
\begin{eqnarray}
  \Phi_{\rm disc}(R,z)=-\frac{G M_{\rm d}} { \sqrt{R^{2} + \left(
        a_d + \sqrt{z^{2}+b_d^{2}} \right)^{2}}},
\end{eqnarray}
with $M_{\rm d}=10^{11}~M_{\odot}$, $a_d=6.5$ kpc and $b_d=0.26$
kpc. Finally, the bulge is modelled as a Hernquist potential
\begin{eqnarray}
  \Phi_{\rm bulge}(r)=-\frac{G M_{\rm b}} {r+c},
\end{eqnarray}
using $M_{\rm b}=3.4\times10^{10}~M_{\odot}$ and $c=0.7$ kpc.  The
superposition of these components gives quite a good representation of
the Milky Way. The circular speed at the solar radius is $\sim 220$ km
s$^{-1}$. $P_{LSR}=2.18\times10^8$ years is the orbital period of the
LSR and $V_c=222.5$ km s$^{-1}$ denotes the circular rotational
velocity at the solar Galactocentric distance, $R_0=8$ kpc.

For our analysis of gradients, we are interested in the mean radial
Galactocentric distance ($R_m$) as a function of the stellar
population and the orbital shape. \cite{Wilson11} has analysed the
radial orbital eccentricities of a RAVE sample of thick disc stars, to
test thick disc formation models. Here we focus on possible local
gradients, so instead consider the {\it vertical} orbital
eccentricity, $e_V$. $R_m$ is defined as the arithmetic mean of the
final perigalactic ($R_p$) and apogalactic ($R_a$) distances, whereas
$e_V$ is defined as follows:

\begin{eqnarray}
e_v=\frac{(|Z_{max}|+|Z_{min}|)}{R_m},
\end{eqnarray}
where $R_m=(R_a+R_p)/2$ \citep{Pauli05}. Due to $z$-excursions
$R_p$ and $R_a$ can vary, however this variation is not more than
5 per cent.

\section{Population Analysis}
\subsection{Classification using space motions}
The procedure of \cite*{Bensby03, Bensby05} was used to separate sample
stars into different populations. This kinematic methodology assumes that
Galactic space velocities for the thin disc ($D$), thick disc ($TD$),
and stellar halo ($H$) with respect to the LSR have Gaussian
distributions as follows:

\begin{equation}
f(U,~V,~W)=k~\times~\exp\Biggl(-\frac{U_{LSR}^{2}}{2\sigma_{U{_{LSR}}}^{2}}-\frac{(V_{LSR}-V_{asym})
^{2}}{2\sigma_{V{_{LSR}}}^{2}}-\frac{W_{LSR}^{2}}{2\sigma_{W{_{LSR}}}^{2}}\Biggr),
\end{equation}
where
\begin{equation}
k=\frac{1}{(2\pi)^{3/2}\sigma_{U{_{LSR}}}\sigma_{V{_{LSR}}}\sigma_{W{_{LSR}}}},
\end{equation}
normalizes the expression. For consistency with other analyses
$\sigma_{U{_{LSR}}}$, $\sigma_{V{_{LSR}}}$ and $\sigma_{W{_{LSR}}}$
were adopted as the characteristic velocity dispersions: 35, 20 and 16
km s$^{-1}$ for thin disc ($D$); 67, 38 and 35 km s$^{-1}$ for thick
disc ($TD$); 160, 90 and 90 km s$^{-1}$ for halo ($H$), respectively
\citep{Bensby03}. $V_{asym}$ is the asymmetric drift: -15, -46 and
-220 km s$^{-1}$ for thin disc, thick disc and halo, respectively. LSR
velocities were taken from \cite{Coskunoglu11} and these values are
$(U, V, W)_{LSR}=(8.83\pm0.24, 14.19\pm0.34, 6.57\pm0.21)$ km
s$^{-1}$.

The probability of a star of being ``a member'' of a given population
is defined as the ratio of the $f(U, V, W)$ distribution functions
times the ratio of the local space densities for two
populations. Thus,

\begin{equation}
TD/D=\frac{X_{TD}}{X_{D}}\times\frac{f_{TD}}{f_{D}}~~~~~~~~~~TD/H=\frac{X_{TD}}{X_{H}}\times\frac{f_{TD}}{f_{H}},
\end{equation}
are the probabilities for a star being classified as a thick disc star
relative to it being a thin disc star, and relative to it being a halo
star, respectively. $X_{D}$, $X_{TD}$ and $X_{H}$ are the local space
densities for thin disc, thick disc and halo, i.e. 0.94, 0.06, and
0.0015, respectively \citep*{Robin96, Buser99}. We followed the
argument of \cite{Bensby05} and separated the sample stars into four
categories: $TD/D\leq 0.1$ (high probability thin disc stars),
$0.1<TD/D\leq 1$ (low probability thin disc stars), $1<TD/D\leq 10$
(low probability thick disc stars) and $TD/D>10$ (high probability
thick disc stars). Fig. 1 shows the $U-V$ and $W-V$ diagrams as a
function of spectral types and population types defined by using
\cite{Bensby03}'s criteria. It is evident from Fig. 1 that the
kinematic population assignments are strongly affected by space-motion
uncertainties.

Using Fig. 1, 15270 and 1142 stars of the sample were classified
as high and low probability thin disc stars, respectively, whereas 287
and 309 stars are low and high probability thick disc stars (Table
1). The relative number of high probability thick disc and thin disc
stars (2 per cent) is evidently very much lower than the number
expected, especially for a sample biased to intermediate Galactic
latitudes.

\begin{figure*}
\begin{center}
\includegraphics[scale=0.70, angle=0]{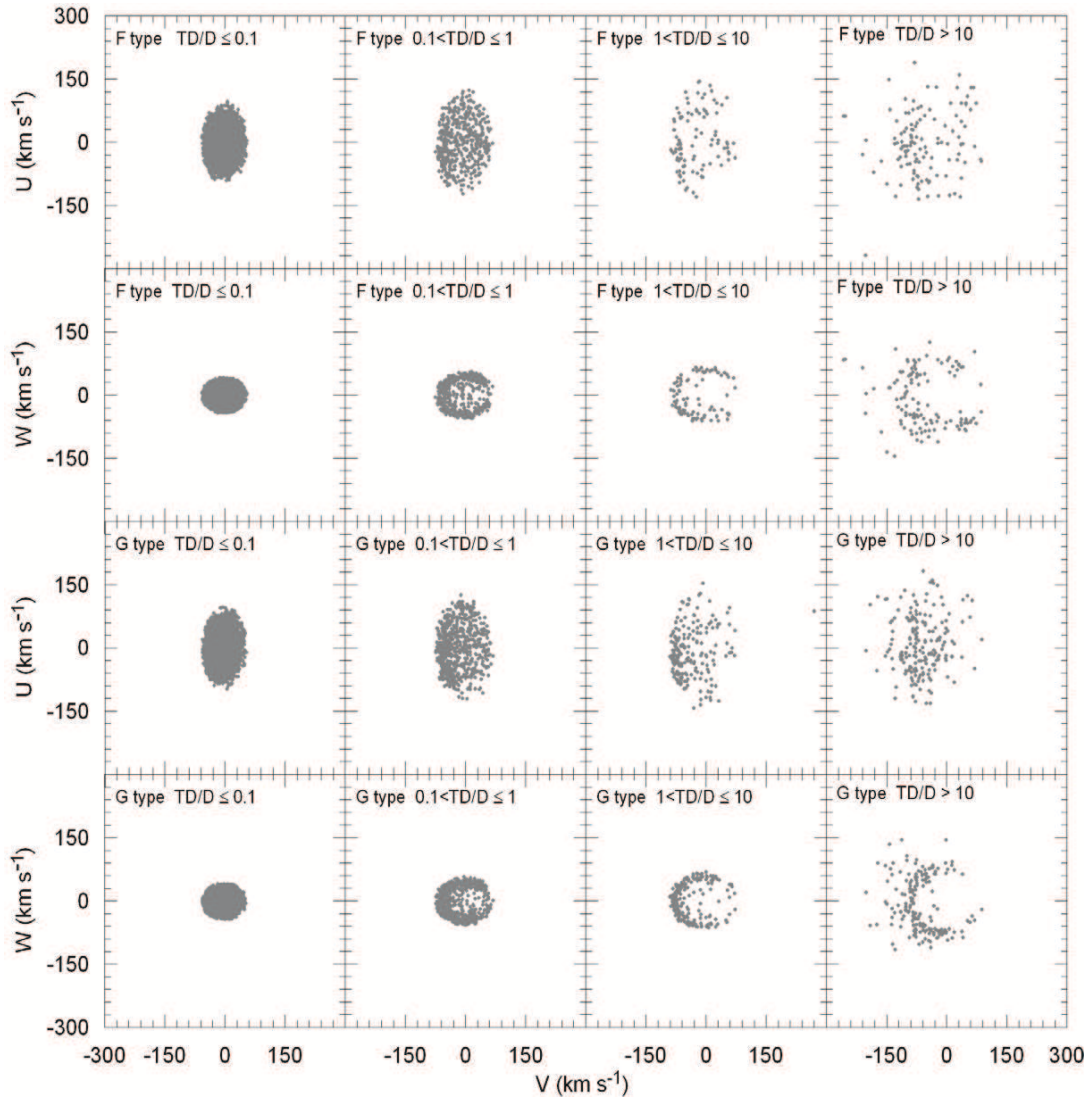}
\caption[] {$U-V$ and $W-V$ diagrams of F- and G-type stars
  applying \cite{Bensby03}'s population classification criteria. It
  is apparent that space motion uncertainties remain significant, even
for this nearby sample.}
\end{center}
\end{figure*}

\begin{table}
\center
\caption{Population frequency distribution of sample stars according to Bensby et al. (2003)'s 
kinematical criteria. $TD/D$ ranges are explained in the text.}
\begin{tabular}{lccccc}
\hline
     &  \multicolumn{5}{c}{Number of stars}\\
\hline
 Spectral Type    & $TD/D\leq0.1$ & $0.1<TD/D\leq1$ & $1<TD/D\leq10$ & $TD/D>10$ & Total\\
\hline
F \& G type stars &  15270 & 1142 & 287 & 309 & 17008\\
F type stars &  9566  & 484  & 97  & 124 & 10271\\
G type stars &  5704  & 658  & 190 & 185 & 6737\\
\hline
\end{tabular}
\end{table}

\subsection{Classification using stellar vertical orbital shape}

\begin{figure}
\begin{center}
\includegraphics[scale=0.50, angle=0]{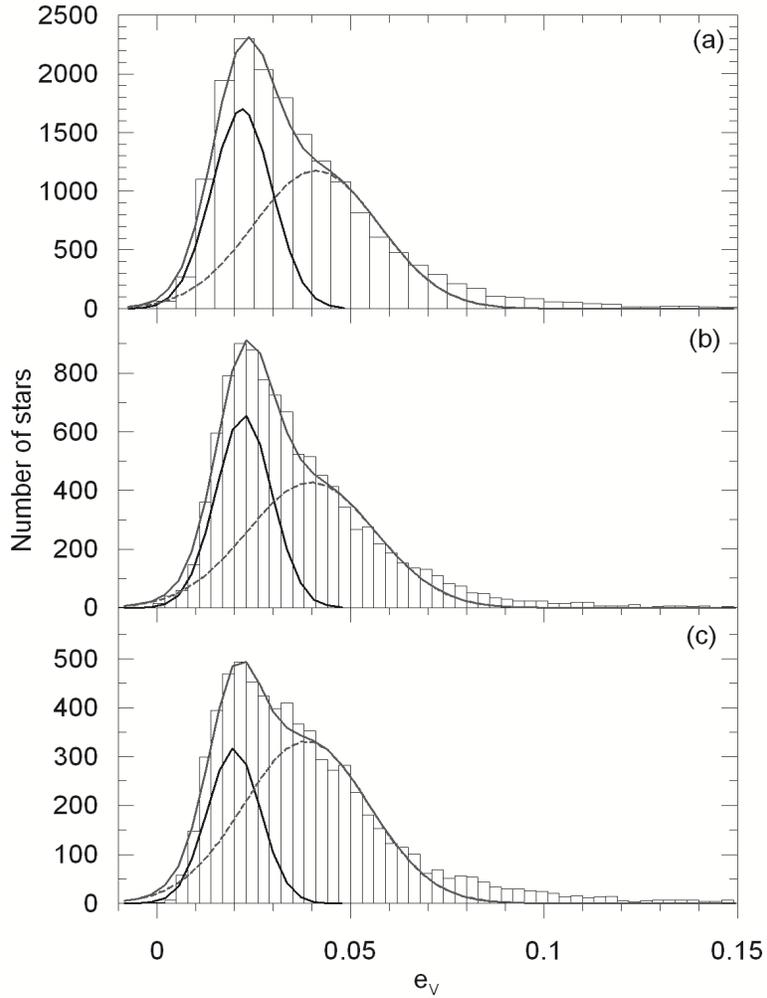}
\caption[] {Three vertical eccentricity distributions: (a) for F- and
  G- type stars, (b) for F-type stars and (c) for G-type stars.}
\end{center}
\end{figure}

Both radial and vertical orbital eccentricities contain valuable
information: here we consider the vertical orbit shape.  Vertical
orbital eccentricities were calculated, as described above, from
numerically-integrated orbits. We term this the dynamical method.
Fig. 2 shows that the distribution function of $e_V$ for F- and G-type
stars, considered separately or in total, are not consistent with
being a single Gaussian distribution. A two-Gaussian model however
does provide an acceptable fit. Hence, we separated our sample into
three categories, i.e. stars with $e_v\leq0.04$, $0.04<e_v\leq0.1$ and
$e_v>0.1$, for each spectral type and fitted their metallicities to
their mean radial distances ($R_m$) in order to investigate the
presence of a metallicity gradient for RAVE stars. We provide 
Table 2 electronically, which includes stellar parameters from RAVE DR3, 
calculated kinematical and dynamical parameters and stellar population 
for the entire sample. The categories and the number of stars 
corresponding to those categories are given in Table 3.    

\begin{landscape}
\begin{table*}
\setlength{\tabcolsep}{4pt}
{\scriptsize
\center
\caption{Stellar atmospheric parameters, astrometric, kinematic and dynamic data for the whole sample: (1): ID, (2): Designation, (3-4): Equatorial coordinates in degrees, (5): $T_{eff}$ in K, (6) $\log g/(cm s^2)$, (7): Calibrated metallicity $[M/H]$ (dex), (8-9): Proper motion components in mas/yr, (10): $d$ in kpc, (11): Heliocentric radial velocity in km s$^{-1}$, (12-17): Galactic space velocity components, and their respective errors in km s$^{-1}$, (18-19): Perigalactic and apogalactic distance in kpc, (20-21): Minimum and maximum distance from the Galactic plane in kpc, (22): $TD/D$ ratio as mentioned in the text.}
\begin{tabular}{lccccccccccccccccccccc}
\hline
(1)	&(2)	&(3)	&(4)	&(5)	&(6)	&(7)	&(8)	&(9)	&(10)	&(11)	&(12)	&(13)	&(14)	&(15)	&(16)	&(17)	&(18)	&(19)	&(20)	&(21)	&(22)\\
ID	& Designation	& $\alpha$ & $\delta$	& $T_{eff}$	& $\log g$	& $[M/H]$	& $\mu_\alpha \cos \delta$	& $\mu_\delta$	& $d$	& $\gamma$	& $U_{LSR}$	& $U_{err}$	& $V_{LSR}$	& $V_{err}$	& $W_{LSR}$	& $W_{err}$ & $d_{per}$	& $d_{apo}$	& $z_{min}$	& $z_{max}$	& $TD/D$\\
\hline
1	&J000018.6-094053	&0.077375	&-9.681528	&6306	&4.16	&0.10	&-1.60	&-7.50	&0.345	&-5.4	&13.34	&3.40	&3.03	&3.32	&8.27	&1.75	&7.678	&8.475	&-0.351	&0.351	&0.006\\
2	&J000020.3-520757	&0.084458	&-52.132528	&5761	&4.06	&0.06	&-12.00	&5.10	&0.289	&30.3	&33.37	&6.13	&18.20	&6.11	&-20.21	&2.97	&7.474	&9.708	&-0.386	&0.394	&0.018\\
3	&J000022.6-265716	&0.094250	&-26.954694	&6080	&4.50	&0.49	&-6.10	&2.60	&0.498	&12.6	&19.15	&10.99	&26.83	&11.10	&-3.00	&3.75	&7.799	&10.091	&-0.566	&0.564	&0.012\\
4	&J000025.9-373329	&0.107750	&-37.558056	&6590	&4.74	&-0.08	&-6.70	&5.20	&0.380	&23.1	&21.68	&2.85	&25.91	&3.55	&-15.08	&1.40	&7.747	&10.007	&-0.473	&0.474	&0.016\\
5	&J000026.3-394148	&0.109500	&-39.696861	&6693	&4.19	&-0.11	&-21.20	&-11.80	&0.259	&21.8	&43.75	&3.26	&10.29	&1.72	&-6.41	&1.20	&7.154	&9.591	&-0.286	&0.285	&0.011\\
6	&J000026.4-671654	&0.110125	&-67.281917	&5773	&4.46	&0.06	&-15.50	&-12.10	&0.158	&26.1	&34.14	&1.95	&-0.85	&1.61	&-5.35	&2.03	&7.023	&8.804	&-0.135	&0.135	&0.007\\
7	&J000029.4-402006	&0.122500	&-40.335083	&5891	&4.16	&-0.16	&-2.00	&-13.10	&0.133	&39.6	&24.15	&0.97	&2.75	&1.09	&-29.43	&0.74	&7.410	&8.719	&-0.367	&0.375	&0.022\\
8	&J000031.4-592708	&0.130667	&-59.452250	&6342	&4.53	&0.22	&8.50	&-13.10	&0.334	&6.7	&9.15	&7.67	&-11.75	&7.94	&8.91	&5.03	&6.986	&7.958	&-0.302	&0.304	&0.007\\
9	&J000033.7-483342	&0.140542	&-48.561889	&6489	&4.10	&-0.06	&-20.10	&-2.60	&0.345	&23.0	&48.53	&4.43	&19.22	&3.65	&-6.71	&2.31	&7.201	&10.201	&-0.374	&0.374	&0.016\\
10	&J000034.3-372147	&0.143042	&-37.363250	&5893	&4.10	&-0.22	&-17.40	&-5.20	&0.121	&-9.3	&16.48	&1.22	&16.17	&0.62	&17.88	&1.01	&7.794	&9.218	&-0.254	&0.254	&0.012\\
11	&J000039.2-505940	&0.163500	&-50.994639	&5746	&4.06	&0.16	&17.20	&0.50	&0.462	&-1.6	&-21.80	&11.22	&-1.94	&11.08	&0.09	&5.21	&7.341	&8.278	&-0.429	&0.429	&0.006\\
12	&J000042.6-495123	&0.177333	&-49.856556	&5771	&4.20	&0.01	&41.20	&-43.50	&0.236	&-1.3	&-13.11	&3.14	&-48.33	&6.53	&16.61	&1.93	&5.115	&7.944	&-0.300	&0.298	&0.062\\
13	&J000042.8-254548	&0.178542	&-25.763528	&6742	&4.67	&0.53	&-28.00	&-14.00	&0.511	&-4.3	&81.15	&9.91	&13.38	&5.39	&22.78	&2.43	&6.586	&11.286	&-0.732	&0.732	&0.102\\
14	&J000100.5-363932	&0.251917	&-36.659028	&5647	&4.52	&0.07	&-11.80	&-17.00	&0.348	&18.8	&42.51	&8.05	&-3.62	&7.91	&-3.71	&2.33	&6.793	&8.983	&-0.364	&0.364	&0.009\\
15	&J000105.4-540845	&0.272333	&-54.146083	&5720	&4.05	&0.05	&16.10	&-0.80	&0.248	&6.5	&-3.13	&3.78	&2.36	&3.59	&-2.53	&2.02	&7.909	&7.973	&-0.219	&0.219	&0.005\\
16	&J000107.9-412207	&0.282917	&-41.368861	&6461	&4.49	&0.31	&-0.60	&4.70	&0.342	&13.6	&11.31	&2.36	&19.14	&2.51	&-8.10	&2.86	&7.830	&9.302	&-0.369	&0.370	&0.008\\
17	&J000108.8-493742	&0.286542	&-49.628556	&5857	&4.68	&0.07	&-35.10	&-27.40	&0.282	&-1.4	&64.18	&6.12	&3.55	&3.18	&30.23	&3.00	&6.592	&10.002	&-0.530	&0.530	&0.077\\
18	&J000124.0-345502	&0.350125	&-34.917417	&5616	&4.85	&0.01	&12.70	&4.50	&0.194	&6.9	&-1.66	&2.15	&12.07	&1.84	&-3.04	&1.09	&7.958	&8.667	&-0.198	&0.198	&0.006\\
19	&J000129.9-595545	&0.374667	&-59.929194	&6413	&4.45	&-0.28	&22.70	&11.80	&0.254	&-31.3	&-28.82	&4.29	&25.28	&3.33	&19.72	&2.45	&7.722	&9.991	&-0.370	&0.370	&0.023\\
20	&J000137.7-701927	&0.407208	&-70.324222	&6158	&4.72	&0.09	&23.50	&-28.90	&0.140	&25.5	&12.37	&1.28	&-20.67	&2.19	&-2.21	&1.34	&6.477	&8.019	&-0.102	&0.102	&0.008\\ 
21	&J000140.8-663745	&0.419917	&-66.629194	&5445	&4.27	&0.25	&49.30	&-9.70	&0.301	&13.6	&-39.63	&9.07	&-34.98	&7.98	&-8.89	&5.26	&5.594	&8.233	&-0.249	&0.249	&0.025\\
22	&J000142.4-290833	&0.426542	&-29.142694	&5696	&4.03	&-0.30	&62.60	&-58.80	&0.173	&-20.0	&-18.33	&2.59	&-53.51	&6.65	&17.43	&1.40	&4.875	&8.013	&-0.273	&0.273	&0.108\\
23	&J000153.2-783310	&0.471792	&-78.552917	&5494	&4.01	&0.44	&12.50	&-2.00	&0.456	&-30.4	&-20.46	&11.51	&17.82	&10.17	&23.40	&10.20	&7.694	&9.256	&-0.460	&0.459	&0.019\\
24	&J000155.8-684858	&0.482292	&-68.816222	&6647	&4.71	&0.46	&10.20	&-12.30	&0.250	&12.9	&10.52	&2.30	&-9.11	&2.33	&4.01	&1.70	&7.102	&8.011	&-0.193	&0.194	&0.006\\
25	&J000159.8-533900	&0.499125	&-53.650056	&6170	&4.28	&0.24	&-0.70	&-10.70	&0.570	&21.6	&32.46	&8.38	&-16.69	&8.43	&0.17	&4.44	&6.403	&8.277	&-0.519	&0.520	&0.009\\
26	&J000203.5-353502	&0.514500	&-35.583917	&6628	&4.96	&0.08	&41.60	&-12.70	&0.244	&-18.3	&-31.05	&4.03	&-20.33	&4.21	&17.06	&1.53	&6.401	&8.252	&-0.331	&0.331	&0.015\\
27	&J000204.0-483953	&0.516875	&-48.665000	&6317	&4.28	&0.25	&-11.30	&-12.60	&0.589	&37.2	&64.83	&7.27	&-11.52	&6.45	&-9.29	&3.54	&6.088	&9.338	&-0.621	&0.622	&0.022\\
28	&J000207.2-582534	&0.530167	&-58.426306	&5624	&4.13	&0.31	&65.80	&-12.60	&0.276	&19.7	&-51.15	&8.28	&-45.87	&6.88	&-18.27	&2.96	&5.035	&8.387	&-0.313	&0.313	&0.107\\
29	&J000207.8-355834	&0.532583	&-35.976222	&6203	&4.29	&0.39	&-1.50	&-24.40	&0.420	&-1.4	&32.20	&6.97	&-28.31	&6.02	&15.25	&2.02	&5.950	&8.255	&-0.468	&0.470	&0.020\\
30	&J000213.6-570608	&0.556583	&-57.102361	&6166	&4.11	&0.41	&26.70	&8.20	&0.537	&-12.2	&-56.57	&13.96	&4.30	&12.63	&-6.09	&7.62	&6.824	&9.564	&-0.519	&0.518	&0.014\\
31	&J000220.6-630157	&0.585792	&-63.032722	&5231	&4.15	&0.12	&-24.40	&8.60	&0.281	&13.0	&42.44	&4.39	&31.34	&4.28	&-4.33	&3.22	&7.451	&10.799	&-0.273	&0.271	&0.023\\
32	&J000222.8-245251	&0.594917	&-24.880861	&6079	&4.48	&0.22	&38.60	&-26.60	&0.241	&-28.7	&-20.74	&2.92	&-36.62	&4.86	&24.80	&1.48	&5.702	&8.048	&-0.404	&0.405	&0.048\\
33	&J000224.1-840559	&0.600292	&-84.099778	&6219	&4.10	&0.04	&7.10	&7.40	&0.295	&-8.5	&0.31	&2.97	&20.79	&2.35	&0.71	&2.71	&7.860	&9.237	&-0.175	&0.175	&0.008\\
34	&J000227.2-081124	&0.613167	&-8.190139	&5672	&4.46	&0.27	&-21.80	&-39.80	&0.285	&-28.5	&57.48	&12.76	&-27.14	&11.04	&20.78	&4.89	&5.757	&8.980	&-0.383	&0.380	&0.051\\
35	&J000228.7-405838	&0.619417	&-40.977444	&6118	&4.09	&-0.25	&85.50	&-2.30	&0.187	&-38.8	&-66.51	&6.78	&-17.49	&3.96	&29.68	&1.66	&6.035	&9.227	&-0.464	&0.464	&0.097\\ 
36	&J000230.7-531342	&0.627917	&-53.228444	&6545	&4.09	&-0.25	&26.00	&-0.60	&0.345	&1.2	&-24.86	&5.24	&-6.74	&4.26	&-2.22	&2.33	&7.090	&8.257	&-0.313	&0.313	&0.006\\
37	&J000232.7-395555	&0.636417	&-39.932139	&5987	&4.06	&-0.10	&25.10	&-2.50	&0.321	&1.1	&-21.95	&5.31	&-7.01	&4.94	&-1.02	&1.71	&7.148	&8.212	&-0.314	&0.313	&0.006\\
38	&J000233.6-531053	&0.639917	&-53.181444	&6064	&4.04	&0.07	&33.00	&-20.10	&0.294	&11.8	&-15.40	&3.96	&-34.19	&5.41	&-0.83	&2.00	&5.793	&7.947	&-0.261	&0.261	&0.014\\
39	&J000246.2-673717	&0.692667	&-67.621611	&5643	&4.28	&0.16	&40.70	&-26.90	&0.185	&1.5	&-13.59	&2.80	&-20.53	&3.44	&13.50	&1.53	&6.531	&7.975	&-0.212	&0.212	&0.010\\
40	&J000255.7-625220	&0.732125	&-62.872250	&5954	&4.61	&-0.07	&-5.10	&-10.80	&0.119	&-19.5	&6.45	&1.28	&18.97	&1.24	&26.08	&1.02	&7.917	&9.267	&-0.348	&0.347	&0.022\\ 
...	&...	&...	&...	&...	&...	&...	&...	&...	&...	&...	&...	&...	&...	&...	&...	&...	&...	&...	&...	&...	&...\\
...	&...	&...	&...	&...	&...	&...	&...	&...	&...	&...	&...	&...	&...	&...	&...	&...	&...	&...	&...	&...	&...\\
...	&...	&...	&...	&...	&...	&...	&...	&...	&...	&...	&...	&...	&...	&...	&...	&...	&...	&...	&...	&...	&...\\
\hline
\end{tabular}
}
\end{table*}
\end{landscape} 

\begin{table}
\center
\caption{Vertical eccentricity frequency distribution of sample stars.}
\begin{tabular}{ccccc}
\hline
              &  \multicolumn{4}{c}{Number of stars}\\
\hline
Spectral Type & $e_v\leq0.04$ & $0.04<e_v\leq0.1$ & $e_v>0.1$ & Total\\
\hline
F type stars  & 6923 & 3145 & 203 & 10271\\
G type stars  & 4223 & 2264 & 250 & 6737\\
\hline
\end{tabular}
\end{table}

\section{Results}
\subsection{Metallicity Gradients from the Kinematical Population
  Assignment Method}

\begin{figure}
\begin{center}
\includegraphics[scale=0.50, angle=0]{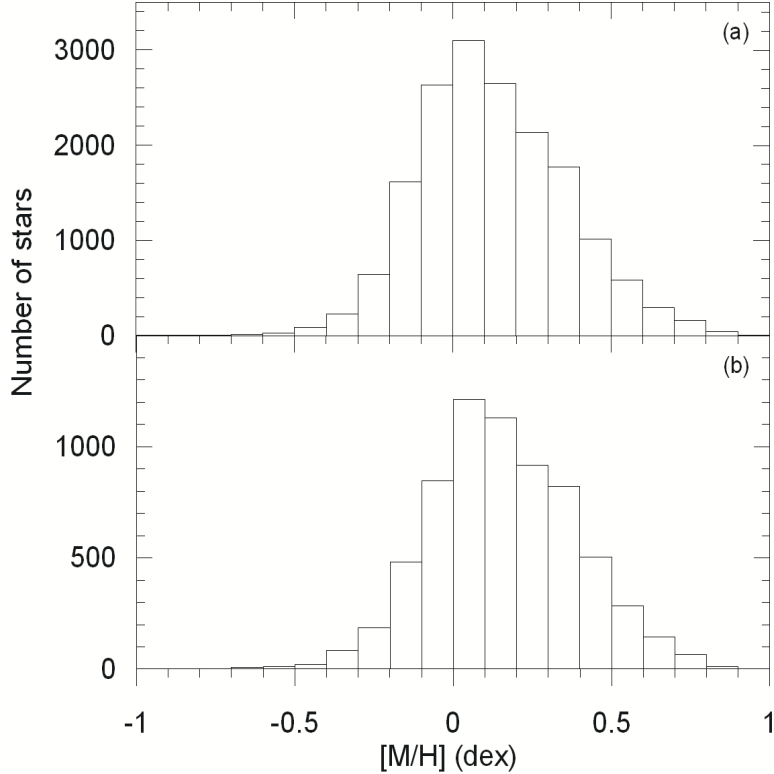}
\caption[] {Metallicity distributions for: (a) F- and (b) G-type stars.}
\end{center}
\end{figure}

The metallicity distribution functions of our final sample, divided
into two spectral types, are shown in Fig. 3. We note these are RAVE
DR3 metallicities $[M/H]$ and not standard $[Fe/H]$ values. As seen from
the figure, the metallicity distribution extends from -0.7 to +1
dex. We now consider the metallicities as a function of the mean
orbital Galactocentric radial distance ($R_m$) for each different
population, i.e. for F and G-type stars and test for the presence of a
metallicity gradient for each population. We fitted the distributions
to linear equations, whose gradient is any metallicity gradient,
$d[M/H]/dR_m$. The results are shown in Fig. 4. The metallicity
gradients in all panels of Fig. 4 are either small or consistent with
zero. The best determined values are for high probability thin disc
stars, where the gradient is $d[M/H]/dR_m$ is $-0.043\pm0.005$ dex
kpc$^{-1}$ for F-type stars, and $-0.033\pm0.007$ dex kpc$^{-1}$ for
G-type stars (Table 4). Interestingly, the metallicity gradient for high 
probability thick disc stars is consistent with zero. However, we treat 
this value with caution, for the number of stars are only 124 and 185 
for F and G-types, respectively, for this population, and Fig. 1 reminds 
us that space motion errors are important. \cite{Wilson11} discusses 
RAVE thick disc stars in more detail.

\begin{figure*}
\begin{center}
\includegraphics[scale=0.70, angle=0]{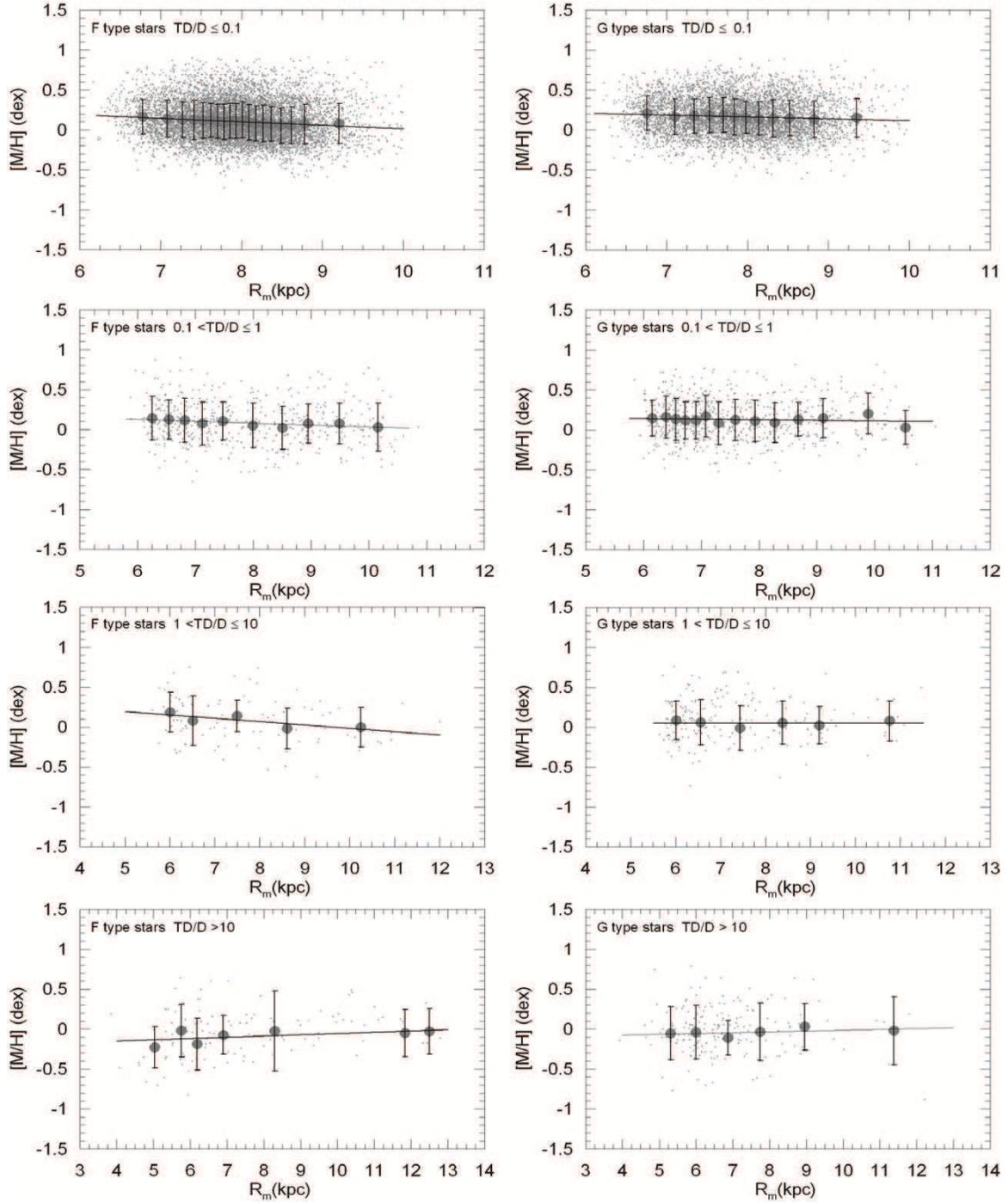}
\caption[] {$R_m-[M/H]$ diagrams for sample stars as a function of spectral types and populations.}
\end{center}
\end{figure*}

\begin{table}
\center
\caption{Radial metallicity gradients for F- and G-type stars
  evaluated from kinematical and dynamical data. Ranges of $TD/D$ and
  $e_v$ are explained in the text.}
\begin{tabular}{lcccc}
\hline
 Population type  & F type stars & Sample size & G type stars & Sample
 size\\
 & $d[M/H]/dR_m$ & & $d[M/H]/dR_m$ & \\
& (dex kpc$^{-1}$) & &  (dex kpc$^{-1}$)\\
\hline
$TD/D\leq 0.1$   & -0.043$\pm$0.005&9566& -0.033$\pm$0.007&5704\\
$0.1<TD/D\leq 1$ & -0.024$\pm$0.007&484& -0.007$\pm$0.009&658\\
$1<TD/D\leq 10$  & -0.042$\pm$0.017&97&  0.000$\pm$0.010&190\\
$TD/D>10$        &  0.016$\pm$0.011&124&  0.010$\pm$0.009&185\\
$e_v\leq0.04$    & -0.051$\pm$0.005&6923& -0.020$\pm$0.006&4223\\
$0.04<e_v\leq0.1$& -0.020$\pm$0.005&3145& -0.004$\pm$0.005&2264\\
$e_v>0.1$        &  0.016$\pm$0.012&203&  0.037$\pm$0.016&250\\
\hline
\end{tabular}
\end{table}

\subsection{Metallicity Gradients from the Dynamical Population
  Assignment Method}

We divide the data into three bins of vertical orbital eccentricity,
and tested for any dependence of the metallicity versus mean radial
Galactocentric distance, $R_m$, again using linear fits.
The results are presented in Fig. 5 and Table 4. We detect a significant
metallicity gradient for high-probability F-type thin disc stars
of $d[M/H]/dR_m=-0.051\pm0.005$ dex kpc$^{-1}$. For G-type stars the metallicity
gradients are rather lower being consistent with zero (Fig. 5 and
Table 4).

These results are consistent with there being somewhat steeper
abundance gradients in younger stars than in older. We test that
further in the next section.

\begin{figure*}
\begin{center}
\includegraphics[scale=0.70, angle=0]{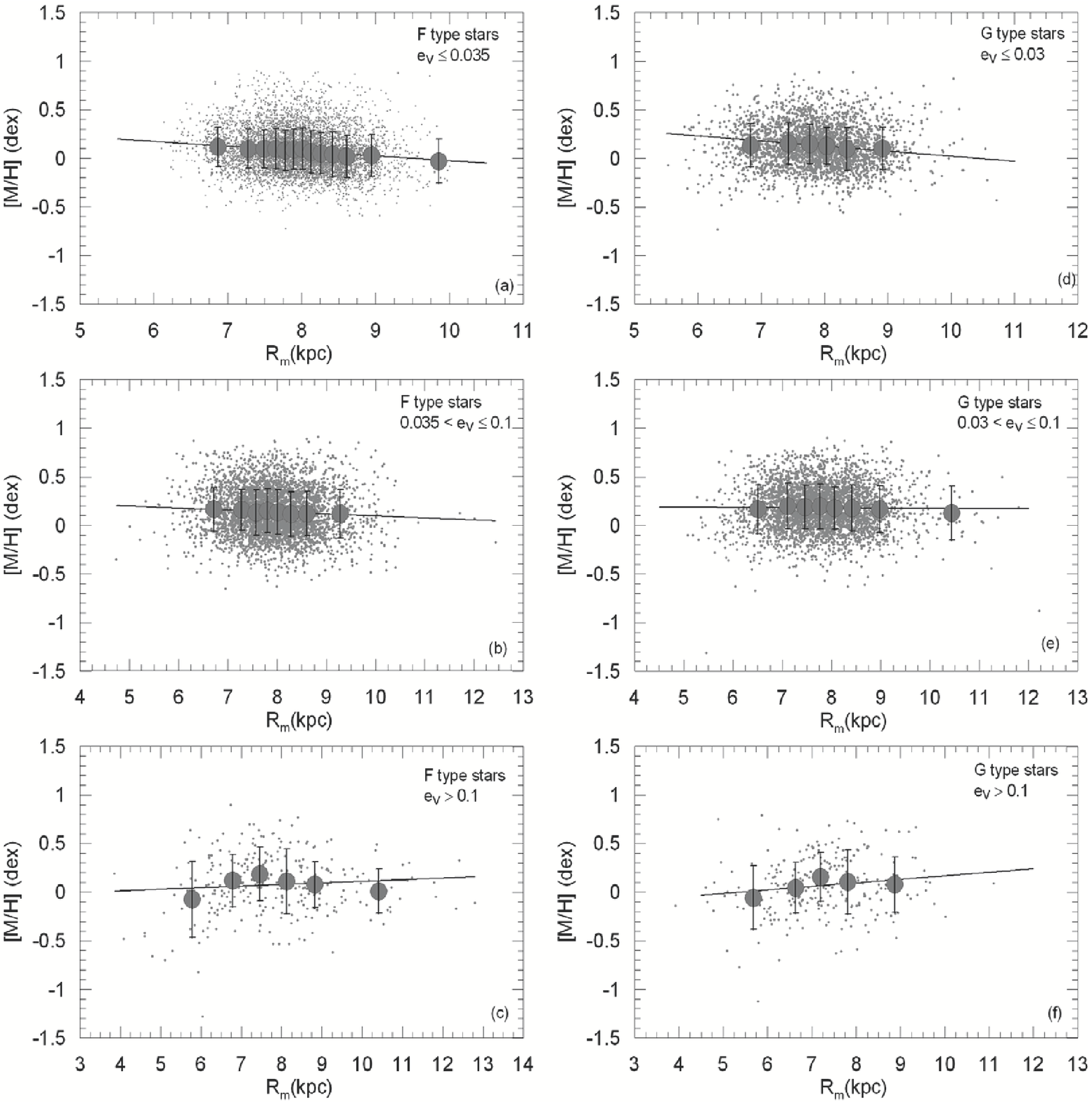}
\caption[] {$R_m-[M/H]$ diagrams for sample stars as a function of
  spectral type and vertical orbital eccentricity.}
\end{center}
\end{figure*}

\subsection{Metallicity Limitation for Metal Rich Stars}

The RAVE pipeline derives metallicity, as any other parameter, using a penalized $\chi^2$ technique by finding an optimal match between the observed spectrum and a spectrum constructed from a library of pre-computed synthetic spectra. The metallicity results match if a similar analysis method is used. The results of the analysis using an independent $\chi^2$ procedure \citep{Munari05} yield metallicities which are entirely consistent with the RAVE pipeline results, i.e. mean offset of $0.04\pm0.02$ dex and a standard deviation of $0.17$ dex. RAVE metallicities as derived from the RAVE pipeline are part of a self consistent native RAVE system of stellar parameters which are tied to a $\chi^2$ analysis using a library of Kurucz template spectra.

However, other spectral methods which derive metallicities from the strengths of individual spectral lines and not from a $\chi^2$ match of synthetic and observed spectra, do not yield very consistent results with those of the RAVE pipeline. As usual, calibration of an internal abundance scale onto standard external systems requires special consideration. One can see the following trends in a comparison of metallicities derived by other methods and the RAVE pipeline: i) The difference between RAVE and reference metallicity increases with an increased $\alpha$-enhancement, in the sense that RAVE values become too metal poor. ii) The difference is larger at lower metallicities. iii) The difference is larger for giants than for main sequence stars, though the variation is much weaker than $\alpha$-enhancement or metallicity. iv) The difference does not depend on temperature.

In \cite{Zwitter08}, the RAVE metallicities ($[m/H]$) derived by the $\chi^2$ method  were calibrated to the metallicities ($[M/H]$) which are in line with the metallicity system of the datasets \em{in situ}\em, such as those appearing in the Asiago Observatory and \cite{Soubiran05} catalogues. The final calibration is as follows:

\begin{equation}
[M/H] = 0.938 [m/H] + 0.767 [\alpha/Fe] - 0.064\log g + 0.404.
\end{equation}
The classical indicator for the metal abundance is the iron abundance, $[Fe/H]$. The following relation between the calibrated metallicity and the iron abundance taken from \cite{Zwitter08} produces a limitation for metal rich stars:   

\begin{equation}
[M/H] = [Fe/H] + 0.11 [1\pm(1-e^{-3.6|[Fe/H]+0.55|})],
\end{equation}
where the plus sign applies for $[Fe/H]<-0.55$ dex and the minus sign otherwise. As the observed upper limit of metal rich stars is about $[Fe/H]=+0.5$ dex, this is the case for calibrated metallicity as well, i.e. approximately $[M/H]\leq0.5$.

\subsection{Metallicity Gradients from Blue Stars with $[M/H]\leq+0.5$}

Our analyses above (Table 4) show that there are differences between
the metallicity gradients estimated from F and G spectral type stars,
in the sense that abundance gradients derived from RAVE F type-stars
are steeper than the metallicity gradients from G type stars. We test
this result further by separating the data into two spectral-type
subsamples, F0-F3 and F4-G9, and estimate metallicity gradients for
these subsamples. The results are shown in Fig. 6. The significant
result is that the metallicity gradients for high probability thin
disc stars ($TD/D\leq$0.1 in Table 4) became steeper,
i.e. $-0.054\pm0.008$ dex kpc$^{-1}$ and $-0.050\pm0.015$ dex
kpc$^{-1}$ for stars of population type F0-F3 and F4-G9 respectively.

As noted in Section 5.3 the RAVE $[M/H]$ distribution does extend to 
very high abundance values, compared to $[Fe/H]$. In order to test if the stars 
with most extreme $[M/H]$ values have systematically unreliable astrophysical 
parameters,we excluded stars with RAVE $[M/H]\geq+0.5$ dex from the sample. 
We then re-estimated metallicity gradients for the two subsamples cited above, 
F0-F3 and F4-G9. The result, given in Fig. 7 confirms our apparent trend, 
in the sense that the metallicity gradient estimated from the bluer stars, 
$-0.065\pm0.018$ dex kpc$^{-1}$, is steeper than that obtained from the later 
spectral type stars, $-0.025\pm0.008$ dex kpc$^{-1}$.

\begin{figure*}
\begin{center}
\includegraphics[scale=0.50, angle=0]{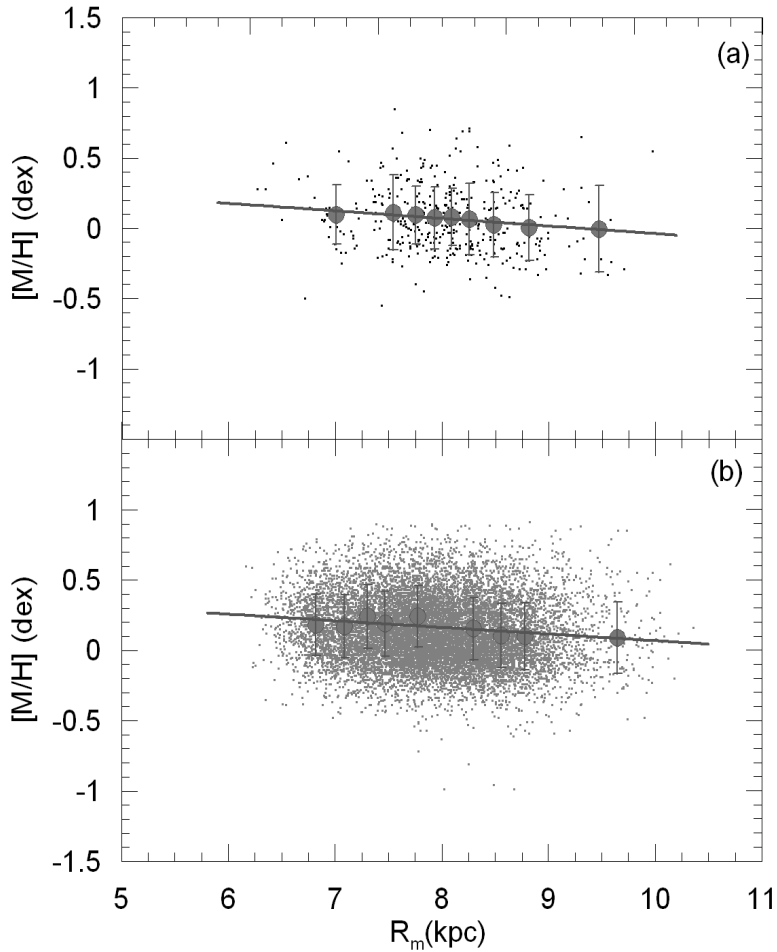}
\caption[] {$R_{m}$-$[M/H]$ diagrams for two subsamples. (a) for F0-F3
 and (b) F4-G9 spectral types.}
\end{center}
\end{figure*}

\begin{figure*}
\begin{center}
\includegraphics[scale=0.50, angle=0]{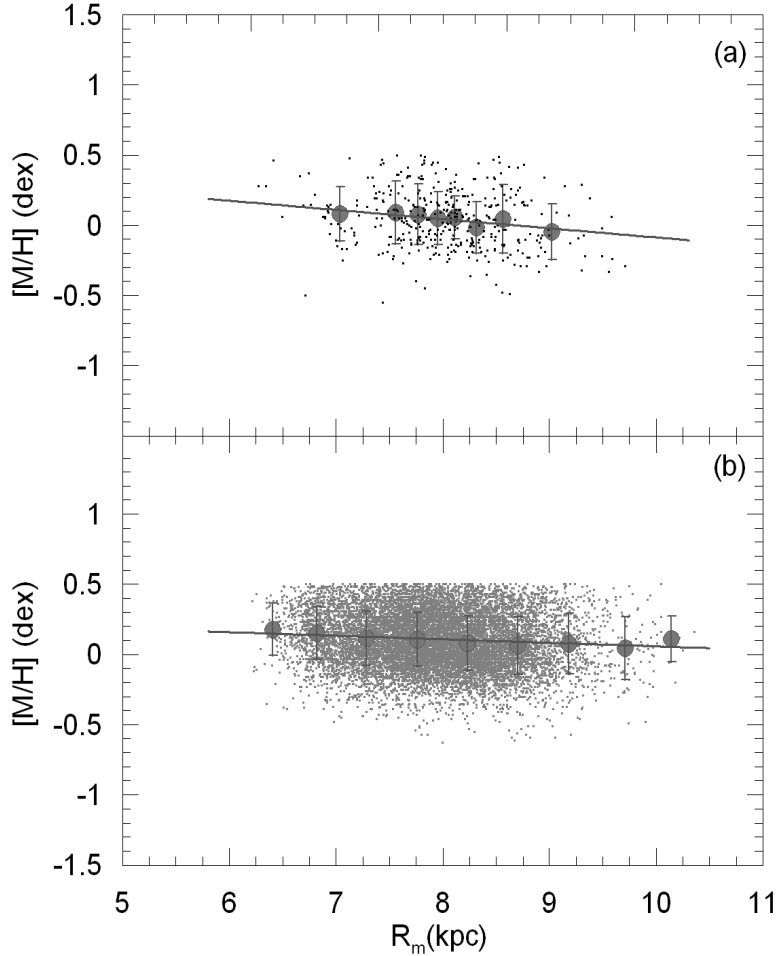}
\caption[] {$R_{m}$-$[M/H]$ diagrams for the same subsamples in Fig. 6, 
but with metallicity restriction with $[M/H]<+0.5$ dex. This selection 
leads to a steeper metallicity gradient for early spectral type stars 
(panel a) and a flatter gradient for later spectral type stars (panel b).}
\end{center}
\end{figure*}

\section{Discussion and Conclusion}

We have used the RAVE DR3 to identify stars classified as
dwarfs, further excluding cool stars, and those with the most
uncertain space motions. We have then obtained a
homogeneous sample of dwarfs defined by: 1) $4<\log g\leq5$,
2) $0.09<(J-H)_0\leq 0.39$, and 3) total space velocity error
$S_{err}\leq 25$ km s$^{-1}$. 10271 stars of this sample are F-type
stars, whereas 6737 of them are G-type stars. For each star we
calculated distances, total space motions, and integrated the stellar
orbit in a Galactic potential to derive both mean Galactocentric
distance and stellar orbital shape. We used the calibrated
RAVE DR3 metallicities and these mean radial distances to investigate the
presence of a radial metallicity gradient, dividing the sample into a
variety of subsamples. These subsamples were defined using
probabilistic population assignment, considering in turn space motions
and stellar Galactic orbital properties - what we term ``kinematic'' and ``dynamic''
methods. Kinematic properties allowed stars to be separated into four
populations, i.e. high probability thin disc stars, low probability
thin disc stars, low probability thick disc stars and high probability
thick disc stars, although only the high probability thin disc
category retained sufficiently large sample size for more detailed
consideration. In the dynamically-defined sample, stars were separated into
three subsamples according to their vertical orbital eccentricity,
i.e. $e_v\leq0.04$, $0.04<e_v\leq0.10$ and $e_v>0.10$.

In all cases, we derive significant metallicity gradients for the
samples which are - statistically - dominated by the youngest thin disc
stars. We derive significant and marginally shallower gradients for
samples which are - statistically - dominated by somewhat older thin disc
stars. We do not detect any gradient in the very small samples of
stars categorised as thick disc. The accuracy and resulting sample
size of our thick disc population assignments is too poor for any
robust statements to be made.

Our radial metallicity gradients for high-probability thin disc stars,
defined using either of our two population classification methods,
($TD/D\leq0.1$ and $e_v\leq0.04$ in Table 4) can be compared to
values in the literature. Our F star gradient is $-0.043\pm0.005$ dex
kpc$^{-1}$. This is somewhat less than the $d[Fe/H]/dR\rm_G =
-0.06\pm0.01$dex kpc$^{-1}$ obtained for young stars 
\citep[Cepheids,][]{Luck11a}, but is consistent with the thin disc radial
metallicity gradient found by \cite{Karatas11} using earlier RAVE
data. \cite{Karatas11} considered RAVE DR2 data, combining all
spectral types, and using the RAVE ``calibrated'' metallicity scale,
to derive $d[M/H]/dR=-0.04$ dex kpc$^{-1}$. Any difference between
the RAVE and Cepheid gradients may be due to calibration differences.
Our thin-disc G star sample, defined using our kinematic approach,
$TD/D\leq0.1$ produces a gradient ($-0.033\pm0.005$ dex kpc$^{-1}$)
which is also consistent with the \cite{Karatas11} value.

When we define our F and G star samples using our orbital shape
(dynamic) definition $e_v\leq0.04$ the resulting abundance gradients
($-0.051\pm0.005$ and $-0.020\pm0.006$ dex kpc$^{-1}$) agree less well
with the \cite{Karatas11} value. Interestingly they are closer to the
value derived from RAVE DR2 data using RAVE DR2 calibrated $[M/H]$
($-0.04$ dex kpc$^{-1}$) than the uncalibrated $[M/H]$ ($-0.07$ dex
kpc$^{-1}$).  The difference between the values derived from RAVE
calibrated $[M/H]$ is partly due to improved statistics, RAVE DR3
having $\approx$1.5 times more stars than does DR2, but also warns
that RAVE metallicities remain to be robustly calibrated onto an
$[Fe/H]$ metallicity scale.

The radial iron gradient for thin disc stars ($4<age<6$ Gyr)
found by \cite{Nordstrom04} is $-0.099\pm0.011$ dex kpc$^{-1}$.  The
chemical evolution model of \cite{Schonrich09} was tuned to provide an
excellent fit to this gradient. In the model the gradient is caused by
radial migration of stars and flow of gas through the disc. This
gradient is at least a factor of two steeper than our values, presented in Table
3. However, as discussed in Section 3, our sample of RAVE stars are
further from the Galactic plane than the \cite{Nordstrom04} stars, so
vertical abundance gradients may - or may not - be an additional
parameter. At face value the amplitude of time-dependent effects in
models such as that of \cite{Schonrich09} may need to be re-tuned to have a 
smaller effect for stars further from the plane than \cite{Nordstrom04}'s sample 
because radial migration of stars and flow of gas through the disc appears to 
be less effective further from the plane. 

The steepest metallicity gradient which we measure,
i.e. $d[M/H]/dR_m=-0.051\pm0.005$ dex kpc$^{-1}$, corresponds to the
subsample of F-type stars defined using orbital shape, with
eccentricities $e_V\leq0.04$. The number of stars in this interval is
6923 (67 per cent of the whole sample). It is interesting to note that this
value is not only consistent with Cepheid values, but agrees with the value
($-0.050$ dex kpc$^{-1}$) predicted by \cite{Rahimi11} from their
cosmologically simulated galaxy for `intermediate' disc stars ($7<age<10$ Gyr). 
In their model negative radial metallicity gradients are due to inside-out 
formation of the disc. Our stellar sample is however biased to very 
much younger ages than those modelled by \cite{Rahimi11}.

Although our samples are small and uncertainties large, we do not
detect any significant abundance gradient in thick disc stars, using
either definition. Our thick disc radial best-fit metallicity
gradients ($TD/D>10$ and $e_v>0.1$ in Table 4) are however
(marginally significantly) shallower than our gradients for thin disc
stars. The radial iron gradient for thick disc stars ($age>10$ Gyr)
found by \cite{Nordstrom04} is $+0.028\pm0.036$ dex kpc$^{-1}$. Our
values are closer to the $+0.01\pm0.04$ dex kpc$^{-1}$ found by
\cite{Ruchti11} for the metal-poor thick disc. Interestingly these
values agree with chemical evolution models of \cite*{Chiappini97, Chiappini01} 
($+0.01 - 0.03$ dex kpc$^{-1}$). These models decompose
the disc into radial annuli that neither exchange gas nor stars. Thick
disc stars are, on average, further from the Galactic plane than thin
disc stars and so should be relatively independent of radial mixing
(unless radial mixing actually evolves a thin disc star into a thick
disc star as suggested by \cite{Schonrich09}). However, within our
large errors, thick disc stars show no radial gradient at all. This
trend is different to what is seen in the thin disc, suggesting these
stars do not share the same origin as the thin disc.

Metallicity gradients we have estimated from F-type stars are steeper
than the ones obtained from G-type stars, for a given population
(Table 4). Given our sample size, the only robust metallicity
gradients we determine are $-0.043\pm0.005$ dex kpc$^{-1}$ and
$-0.051\pm0.005$ dex kpc$^{-1}$ respectively, for stars classified as
high probability thin disc, i.e. for population types labeled with
$TD/D\leq0.1$ and $e_{V}\leq0.04$. Separation of the sample into two
colour sub-samples, F0-F3 and F4-G9 here, provides a steeper
metallicity gradient, $-0.065\pm0.015$ dex kpc$^{-1}$, for the earlier
spectral type stars. Other subsamples from the RAVE DR3 dwarf sample
are absolutely small in number, and have relatively large
uncertainties. The RAVE DR3 dwarf star sample probes stars with orbits
with mean Galactocentric radii within 3 kpc of the Solar Galactocentric
radius. The stars with spectral types F0-F3 which provide the
steeper metallicity gradient of $-0.065\pm0.015$ dex kpc$^{-1}$,
make up only 3 per cent of the F and G type stars with $[M/H]<0.5$
dex.

From our analysis, we may conclude that the RAVE DR3 data may be
described as two different subsamples, i.e. a thin disc sample
biased to young ages, with a detected metallicity gradient, and a
thin disc sample biased to somewhat older ages, in which we do not
detect any metallicity gradient.

\section{Acknowledgments}
We would to thank the referee Dr. B. Barbuy for her comments and suggestions.
This work has been supported in part by the Scientific and
Technological Research Council (T\"UB\.ITAK) 108T613. G. M. 
Seabroke is funded by the UK VEGA {\it Gaia} Data Flow System grant.

This publication makes use of data products from the Two Micron All
Sky Survey, which is a joint project of the University of
Massachusetts and the Infrared Processing and Analysis
Center/California Institute of Technology, funded by the National
Aeronautics and Space Administration and the National Science
Foundation.  This research has made use of the SIMBAD, NASA's
Astrophysics Data System Bibliographic Services and the NASA/IPAC
ExtraGalactic Database (NED) which is operated by the Jet Propulsion
Laboratory, California Institute of Technology, under contract with
the National Aeronautics and Space Administration.

\end{document}